\documentclass[%
 reprint,
 amsmath,amssymb,
 aps,
showkeys,
]{revtex4-2}

\usepackage{graphicx}
\usepackage{dcolumn}
\usepackage{bm}

\begin{document}

\preprint{APS/123-QED}

\title{Stochastic switching and squeeze of hysteresis window in nonlinear responses of silicon nitride membrane nanoelectromechanical resonators}
\author{Srisaran Venkatachalam}
\affiliation{CNRS, Université Lille, Centrale Lille,Université Polytechnique Hauts-de-France, UMR8520, IEMN, Av. Henri Poincare, Villeneuve d'Ascq 59650, France}
\author{Xin Zhou}
\email{Corresponding Author: xin.zhou@cnrs.fr}
\affiliation{CNRS, Université Lille, Centrale Lille,Université Polytechnique Hauts-de-France, UMR8520, IEMN, Av. Henri Poincare, Villeneuve d'Ascq 59650, France}

\date{\today}

\begin{abstract}
In this work, we present the effects of stochastic force generated by white noise on the nonlinear dynamics of a circular silicon nitride membrane. By tuning the membrane to the Duffing nonlinear region, detected signals switching between low- and high-amplitudes have been observed. They are generated by noise-assisted random jumps between bistable states at room temperature and exhibit high sensitivity to the driving frequency. Through artificially heating different mechanical vibration modes by external input of white noise, the switching rate exhibits exponential dependence on the effective temperature and follows with Kramer’s law. Furthermore, both the measured switching rate and activation energy exhibit sensitivity to the width of the hysteresis window in nonlinear response and the driving force, which is in qualitative agreement with the theoretical descriptions. Besides, white noise-induced hysteresis window squeezing and bifurcation point shifting have also been observed, which are attributed to the stochastic force modulation of the spring constant of the membrane. These studies are essential for the generation and manipulation of stochastic switching effects, paving the way to explore new functions based on probability distributions in nanomechanical resonators.
\end{abstract}

\keywords{stochastic switching, Duffing nonlinearity, silicon nitride membrane, nanoelectromechanical resonator}

\maketitle


\section{\label{sec:intro} Introduction}

Stochastic switching between coexisting states has been observed in chemistry, physics, biology, and engineering systems. Its mechanism can be explained simply through descriptions of a fictive particle hopping randomly in bistable states of a double-well potential  \citep{gammaitoni1998stochastic}. Although these individual switching events are random, both distribution and probability of switching can be well predicted and controlled, exhibiting several potential applications. For instance, it brings an enhancement of signal-to-noise ratio \citep{badzey2005coherent} and random number generators \citep{vodenicarevic2017low}, which could offer a means of enhancing signal processing . Besides,  stochastic switching can directly perform stochastic computing, which are attractive for novel error-tolerant computing schemes in neuromorphic applications \citep{gaba2013stochastic, zahari2020analogue}. Therefore, reserach interests are preserved in exploring stochastic switching in various systems, including Josephson junctions \citep{wiesenfeld1995stochastic, muppalla2018bistability}, protein folding \citep{wales2003energy}, nanomagnets \citep{spano1992experimental, wernsdorfer1997measurements}. 

Micro- or nano-mechanical resonator, which allows electrical and optical signals to couple with a mechanical degree of freedom, is one of interesting components for exploring stochastic switching \citep{stambaugh2006noise,venstra2013stochastic}. Because of intrinsic nonlinearity coming from geomentry, nano-mechanical resonators can be driven as Duffing oscillators, providing the  indispensable condition: bistable states. In addition, the idea of building multifunctional components based on nanoelectromechanical resonators has been driving researchers to explore fundamental aspects and potential applications beyond sensing \citep{maillet2017nonlinear, zhou2019chip, dion2018reservoir, gazizulin2018surface, guerra2010noise}. So far, stochastic switching assisted by white noise has been investigated with cantilevers \citep{venstra2013stochastic}, double-clamped beams \citep{defoort2015scaling, aldridge2005noise}, and membranes \citep{dolleman2019high,chowdhury2017phase}. In very recent years, high-stress silicon nitride membrane electromechanical resonators emerge and offer high quality factors with resonance frequency in MHz ranges. They allow vibrating membranes to be strongly coupled to external electronic circuits, providing a highly sensitive actuation and detection scheme \citep{zhou2021high, pokharel2022capacitively, yuan2015silicon}. However, up to now, there are still few studies on the effect of stochastic forces generated by white noise on the nonlinear behaviors of such membrane electromechanical resonators, such as stochastic switching and hysteresis \citep{dolleman2019high}. 

In this work, we present experimental studies of stochastic switching in a circular membrane nanoelectromechanical resonator, which consists of a silicon nitride drum, capacitively coupled to a suspended aluminum gate. At the room temperature experiment, the membrane resonator is frequency biased in Duffing nonlinear region, creating bi-stable states for mechanical displacements. The membrane is artificially heated by adding external white noise, and the switching rate of different mechanical modes as a function of the effective temperature has been measured. Its characteristics follow Kramer's law. Experimental parameters affect on this noise assisted jumping between bistable states have been analyzed by comparing our measurements with the theoretical descriptions. Besides, squeeze of hysteresis window in nonlinear responses has also been observed when the amplitudes of stochastic force increases. 	

\section{\label{sec:level1} methods}
\begin{figure*}
  \includegraphics[width=0.98\textwidth]{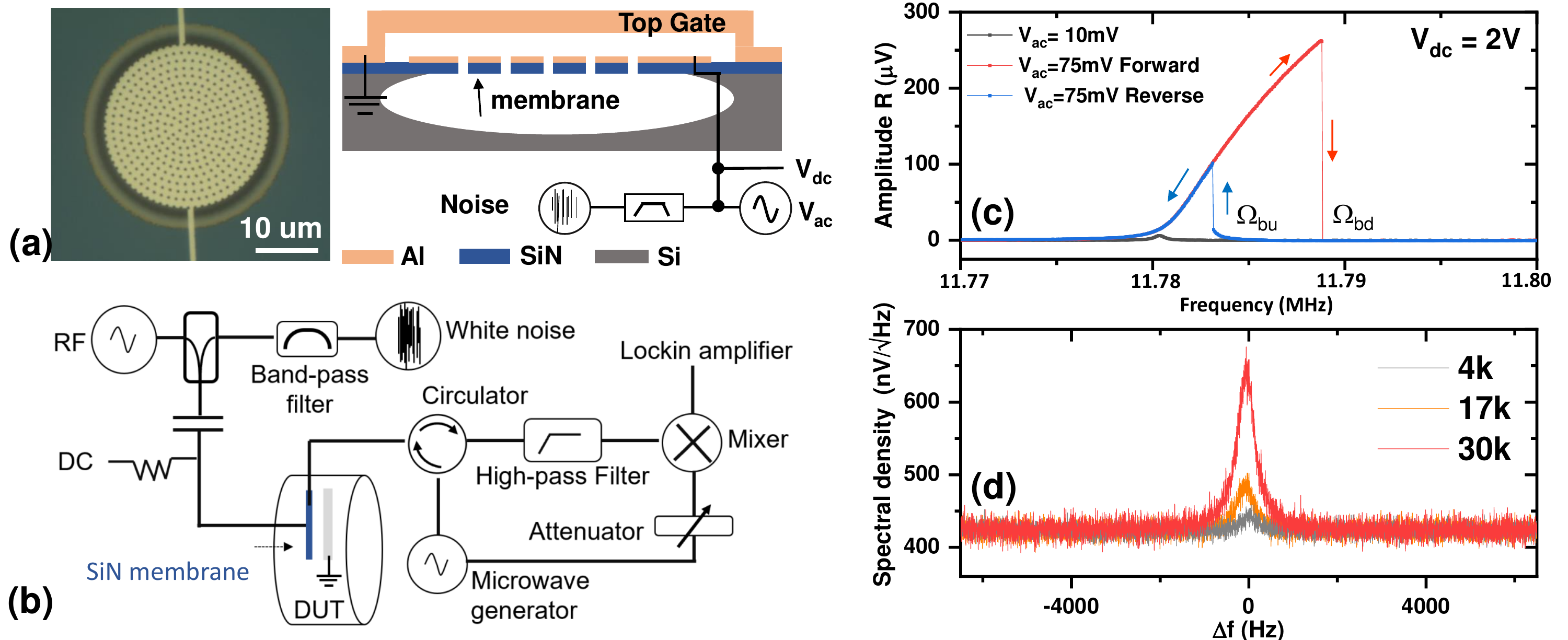}
  \caption{(a) Left, an optical image of a silicon nitride membrane, covered with an Al thin film $\sim$25 nm in thickness. The silicon nitride (SiN) membrane is released from Si substrate by reactive ion etching un-wanted silicon nitride parts,  circular holes ($\sim$ 300 nm in diameter), followed with XeF$_2$ selective etching of the Si layer below the silicon nitride layer through those holes. Right, schematic diagram of the device in lateral view. Note that SiN is silicon nitride. (b) The measurement setup. A silicon nitride membrane is driven by means of an electrostatic force through combing $dc$ and $ac$ signals, and stochastic forces generated by a Gaussian white noise which is filtered to have a band-pass around resonance frequency of the desired mechanical mode. All measurements are performed at room temperature, in vacuum (pressure $< 10^{-6}$ mbar). (c) Black line, linear response of silicon nitride membrane, measured with $V_{dc}$ = 2 V, $V_{ac}$ = 1 mV$_{p}$ for the fundamental mode $\Omega_{01}$. Both blue and red lines show the Duffing responses, which are measured with the higher driving amplitude, $V_{dc}$ = 2 V, $V_{ac}$ = 75 mV$_{p}$. The hysteresis regime is formed between forward (red line) and the backward (blue line) frequency sweeps. (d) Spectral density of the membrane's displacements, at different effective temperature $T_e$.} 
  \label{sch:device_basic}
\end{figure*}
The device measured in this experiment is a silicon nitride membrane electromechanical resonator. It consists of a silicon nitride circular membrane with $\sim$ 33 $\mu$m in diameter and $\sim$ 80 nm in thickness, which is fabricated based on a substrate composed of a high tensile stress ($\sim$ 1 GPa) silicon nitride thin film on top of the silicon wafer. The membrane is covered with an aluminum thin film, $\sim 25$ nm in thickness, and is capacitively coupled to a suspended aluminium top-gate, as shown in Fig.\ref{sch:device_basic} (a). The distance between the membrane and its gate is $d \approx$ 600 nm. Details of the fabrication process have been reported in our previous work \cite{zhou2021high, pokharel2022capacitively}. In this room temperature measurement, the membrane is set in a vacuum chamber ($\sim$ 10$^{-6}$ mbar) and grounded through its top gate. In this capacitive coupling scheme, mechanical displacement $x$ can be excited by an electrostatic force $f_d(t)$, which is generated through combining a $dc$ signal $V_{dc}$ and a $ac$ signal with a frequency $\Omega$, $V_{ac} \cdot cos(\Omega \cdot t)$, which gives  $f_d$ = $\frac{1}{2}\partial(C_g(x)(V_{dc}+V_{ac}))^2/\partial x$. The $C_{g}(x)$ is the coupling capacitance between the membrane and its gate, and the $x(t)$ is the mechanical displacement. To study noise assisted random jumps between bistable states, white noise that is generated by an arbitrary waveform generator is injected to the membrane resonator, through a band-pass filter centred at the desirable frequency corresponding to the resonant mechanical mode. It yields the stochastic force $f_n(t)$ also driving on the membrane. Therefore, the mechanical displacement $x(t)$ of this circular membrane can be described by a motion equation Eq.\ref{eqn:motionEq}, 
\begin{equation}
m_{eff} (\ddot{x}+ \gamma_{m}\dot{x} +\Omega_{m}^2 x + \alpha \, x^3 ) = f_n(t) +  f_d(t).
  \label{eqn:motionEq}
\end{equation}
Here, $\Omega_{m}$ is the resonance frequency of mechanical resonator, the $\gamma_{m}$ is the linewidth, $\alpha$ is the Duffing coefficient, and $m_{eff}$ is effective mass of the membrane depending on the mechanical vibrating modes. 

A detection scheme is built on microwave interferometry, in which the tiny mechanical displacements excited by $f_d[\Omega]$ are transduced by a microwave signal with frequency $\omega$ to be the signal having frequency at $\omega + \Omega$. The displacement $x(t)$ is readout by a lock-in amplifier through a frequency down-conversion. The details of both driving and detection schemes are shown in  \ref{sch:device_basic} (b). The microwave interferometry allows to transduce the detected amplitude of electrical signals into mechanical displacement $x$ by using the relation $V_{out} = G \omega Z_0 C_g x V_{\mu w}/(2d)$ \citep{zhou2021high}. Here, the $Z_0$ = 50 Ohm is the impedance of the measurement line, $C_g \approx$ 10 fF is the capacitance between the membrane and its coupled gate, $V_{\mu w}$ is the input amplitude of microwave signal for detection (in this measurement, $V_{\mu w} \approx$ 300 m$V_p$), the $V_{out}$ is the detected amplitude of the microwave signal, and the $G$ is the total gain in the detection chain in this experiment.

Figure \ref{sch:device_basic} (c) shows linear responses of the mechanical fundamental mode (black line), whose resonance frequency is $\Omega_{m}/2 \pi \approx$ 11.781 MHz with linewidth $\gamma_{m} / 2 \pi$ = 1015 Hz. When the mechanical resonator is driven by a periodic force with a large amplitude, the spring hardening ($\alpha > 0$) makes the resonance frequency shift towards the higher values and frequency responses exhibit hysteric behavior between forward (red line) and backward (blue line) frequency sweeps. Bistable states exist in this region, which is defined between two saddle-node bifurcation points corresponding to the amplitude of mechanical response jump-down $\Omega_{bd}$ in the forward frequency sweep and jump-up $\Omega_{bu}$ in the backward sweep \cite{ aldridge2005noise, venstra2013stochastic, defoort2015scaling}. It is a typical Duffing nonlinear phenomenon and has been observed in various mechanical resonators. Besides, the spectral density of the electromechanical resonator's amplitudes, driven by white noise with different power, has also been measured, as shown in Fig.\ref{sch:device_basic} (d). Based on the basic principle of microwave interferometry mentioned above, the mean square amplitude $<x(t)^2>$ can be obtained from spectral density, which gives the effective temperature of mechanical membrane, $T_e$, from the definition in Eq.\ref{eqn:calT}, 
\begin{equation}
T_{e} = \frac{m_{eff} \Omega_m^2 <x(t)^2>}{k_B}.
  \label{eqn:calT}
\end{equation}
Here, the $k_B$ is the Boltzman constant. The corresponding effective temperatures are marked in Fig. \ref{sch:device_basic} (d), which are obtained based on these expressions by taking parameters of $m_{eff} \approx 4 \times 10^{-14}$ kg and $\omega /(2\pi)$ = 7.01 GHz. We use this method to obtain the value of the $T_e$ when the membrane is artificially heated up by external white noise. In this work, the maximum vlues of the measured spectra density is $\sim$1 nm/$\sqrt{Hz}$.
\section{Experimental results and discussion}
\begin{figure}
  \includegraphics[width=0.48\textwidth]{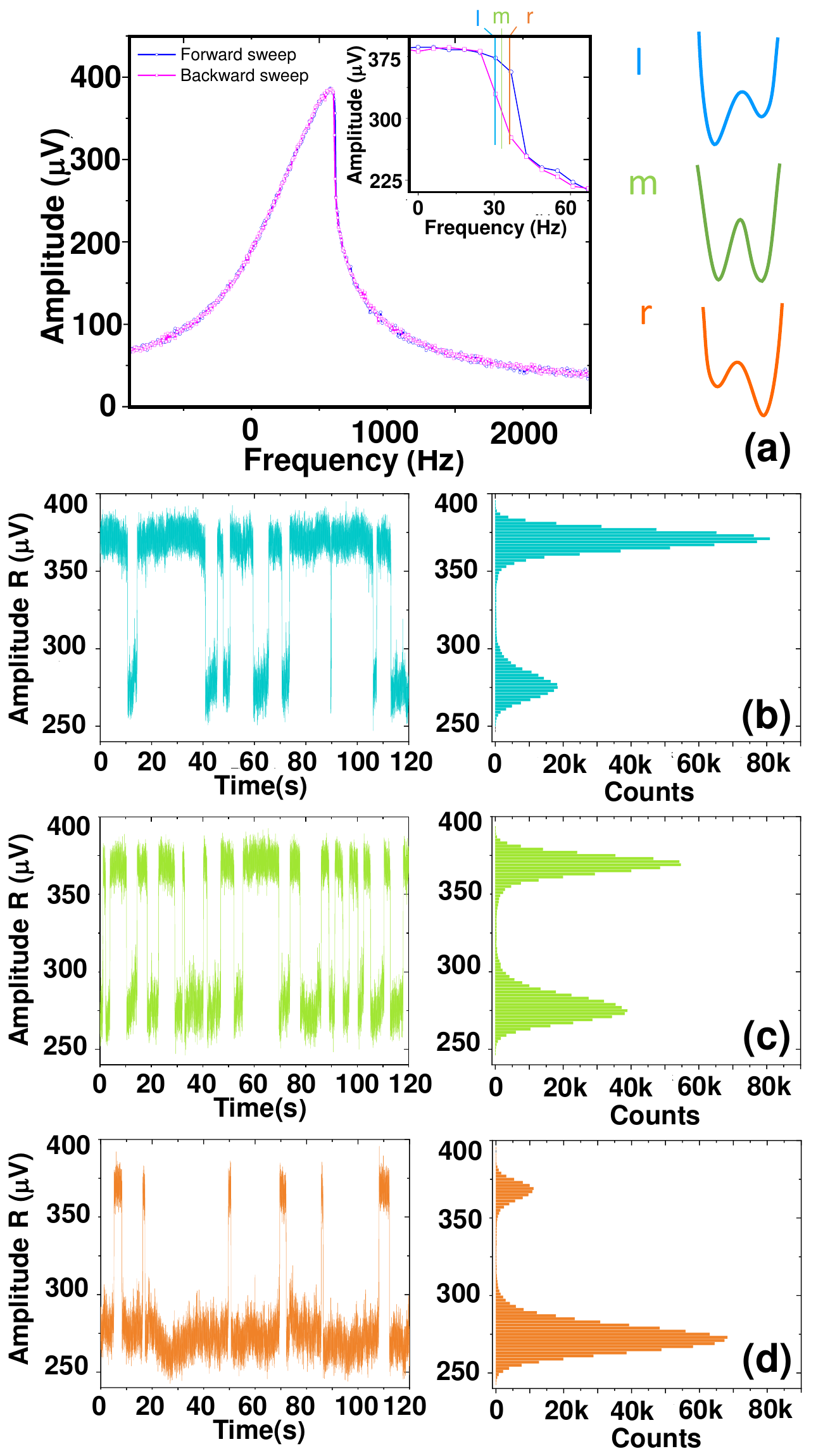}
  \caption{(a) Left, without adding extra white noise, nonlinear responses of the forward and backward frequency sweeps, which are measured with a high driving force generated by $V_{dc}$ = 2 V and $V_{ac}$ = 9 m$V_p$, . Inset, zoom of the hysteresis frequency window, around 15 Hz in width. Right, schematic diagram of double wells corresponding to three different driving frequencies in bistable region, which is marked on the inset figure as "l", "m", and "r" respectively. (b)-(d) The left side of figures show amplitudes as a function of the measurement time in this stochastic system, driving respectively at "l", "m", and "r", as marked in the inset of the (a). Their histograms of the distributions in amplitudes are shown on the right side.}
  \label{sch:switching}
\end{figure}

In order to investigate the switching dynamics of the system in this room temperature measurement, we prepare a small hysteresis window of nonlinear responses to prepare bistable states. In this experiment, it is a challenge to observe the detected signal switching between low- and high- amplitudes in the large hysteresis window, e.g. a width much larger than 100 Hz. It could be due to the mechanical resonance frequency drifting at room temperature. Thus, the width is controlled to be less than 100 Hz by regulating amplitude of the $f_d$, as shown in Fig.\ref{sch:switching} (a). The typical value of the detected mechanical displacement in this nonlinear region is $\sim$ 38 nm. Without injecting extra white noise, stochastic switching between bistable states has been observed by setting the frequency of the driving force in the hysteresis regime. Amplitudes of the detected signals switch between two states, corresponding to the low ("L") and the high ("H") amplitude, as shown in Fig.\ref{sch:switching} (b)-(d). When the driving frequency is biasing at the center of the hysteresis window, the residence of each state exhibits almost equal probability,  as shown in Fig. \ref{sch:switching} (c). Whereas one of the states is more likely to be occupied when the frequency of the driving force is detuned from the center, even for only a few Hz, as presented in Fig. \ref{sch:switching} (b) and (d). The probability of the occupation states exhibits high sensitivity to the driving frequency. This is because the double-well potential, in descriptions of bistable states, has been tilt by the frequency of the periodic driving force, $f_d$. 
\begin{figure}
  \includegraphics[width=0.48\textwidth]{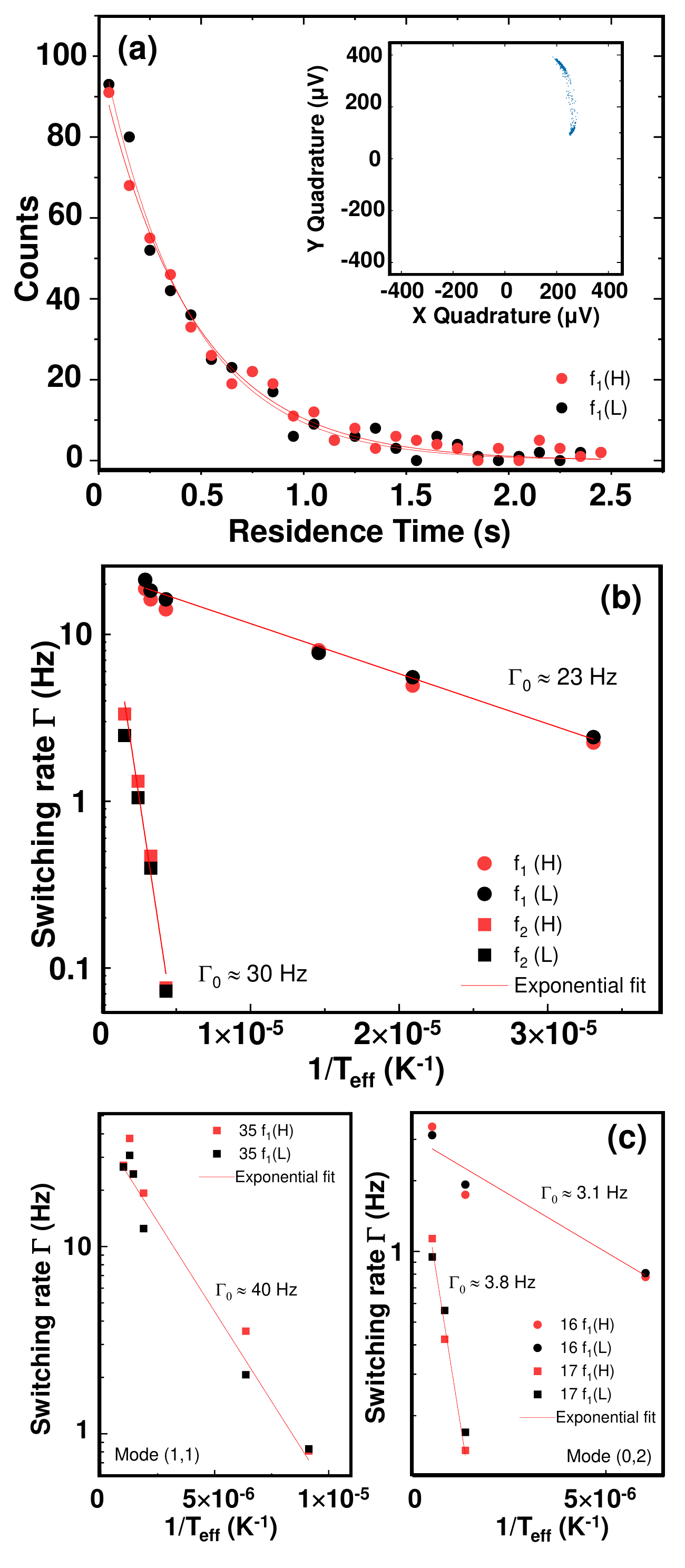}
  \caption{(a) Residence time distributions for both "L" and "H" states at $T_{e}\approx 3 \times 10^4$ K. The red curves are fit by Poisson distribution function, giving a value of switching rate $\Gamma \approx 2.2$ Hz. The inset figure is amplitude distributions of detected signals shown in the two quadrature space and each point is one sample in the measurement of amplitude as a function of time. (b) Switching rate $\Gamma$ of both states as a function of $T_{e}^{-1}$, corresponding to different driving forces. In both (a) and (b), the driving force $f_1$, $f_2$ are generated through combing $V_{dc} $ = 2 V with $V_{ac}$ = 14 m${V_p}$, and 18 m${V_p}$, respectively. (c) Left and Right show $\Gamma$ versus $1/T_e$ for mechanical mode (1,1) and (0,2). These driving forces are normalized with $f_1$.} 
  \label{sch:switchingRateTemp}
\end{figure}

From the previous reports \citep{dykman1979theory}, we know that the switching rate $\Gamma$ in the stochastic switching between the bistable states follows Kramer’s law,
\begin{equation}
  \Gamma = \Gamma_0 \, exp(\frac{-E_a}{k_B T_{e}}).
  \label{eqn:switching}
\end{equation}
The $\Gamma_0$ is the maximum switching rate, $k_B$ is the Boltzmann constant,  $E_a$ is an activation energy corresponding to the barrier height between bistable states, and $T_{e}$ is effective temperature \citep{aldridge2005noise}. For Duffing oscillators, the width of hysteresis window $\vert \Omega_{bd} - \Omega_{bu} \vert$ and the driving frequency $\Omega_d$ are supposed to play important role in manipulating the $E_a$ and the $\Gamma_0$. In the switching process, the distribution of residence time ($\tau$) corresponding to the "L" or "H" state follows a Poisson law, 
\begin{equation}
  N(\tau ) = I \cdot \Gamma \, exp(-\tau \Gamma).
  \label{eqn:poisson}
\end{equation}
Here, the $I$ is a parameter for normalizing the residence time distribution. The probability of residence in each state is therefore sensitive to the driving frequency.

Thus, we first measure the distributions of the residence time for both "L" and "H" occupation states corresponding to each value of the $T_e$. To do so, we exploit two forces to drive the membrane. The periodic driving force $f_d$ with frequency $\Omega_d$ is used  to control the width of hysteresis window and bistable states are prepared by driving membrane at the frequency $\Omega_d \sim (\Omega_{bd}-\Omega_{bu}) /2$. The stochastic force $f_n$ generated by white noise is employed to control the effective temperature $T_e$ of the mechanical mode, by artificially "heating up" the membrane. By measuring the time trace of the detected signal amplitudes, we can make statistics of residence time. Figure \ref{sch:switchingRateTemp} (a) shows distributions of residence time for both "L" and "H" states exhibit equal counts, which are measured at $T_e \approx 3 \times 10^4$ K. The inset figure shows amplitudes of the signal in a two-quadrature space, which presents bistable states of the membrane. These results indicates that potential well is controlled to be symmetric. Then, by using Poisson's law to fit the distribution of residence time, we can obtain the transition rate $\Gamma$ corresponding to different $T_e$. 

Figure \ref{sch:switchingRateTemp} (b) shows exponential plots of $\Gamma$ as functions of $1/T_e$ for the fundamental mode (01), corresponding to different widths of the hysteresis window controlled by the driving force $f_d$. The maximum switching rate $\Gamma_0$ can be obtained by exploiting Kramer’s law as described by Eq.\ref{eqn:switching}. The value of $\Gamma_0 \approx$ 23 Hz is obtained when the width of the hysteresis window $(\Omega_{bd}-\Omega_{bu})/2/\pi$ is controlled to be 18 Hz by the driving force $f_1$. The relatively higher value of $\Gamma_0 \approx$ 30 Hz is obtained by using the higher force $f_2$, yielding the relatively wider hysteresis window with width $(\Omega_{bd}-\Omega_{bu})/2/\pi$ = 67 Hz. Besides, we also measured the switching rate of the higher mechanical modes, (11) and (02), as shown in Fig. \ref{sch:switchingRateTemp} (c). In the measurement of the mode (11), a similar width of hysteresis window, around 62 Hz, has been set and we obtain $\Gamma_0 \approx$ 40 Hz. While, for the mode (02), the hysteresis window is controlled to be quite small, less than 10 Hz, and the measured values of $\Gamma_0$ are $\approx$ 3.1 Hz and 3.8 Hz, well below the damping rate of mechanical resonator, $\gamma_m / (2\pi)$. These measurement results indicate that the maximum value of switching rate between bistable states relies on the hysteresis window $(\Omega_{bd}-\Omega_{bu})$, in accordance with theoretical descriptions \citep{dykman1979theory, defoort2015scaling}. Besides, we also found that the slope of these plots is sensitive to the amplitude of the driving force $f_d$, as shown in the Fig.\ref{sch:switchingRateTemp}(b)-(c). From Kramer's law, we know that these slopes are proportional to the activation energy $E_a$. Therefore, these measurement results demonstrate that the activation energy $E_a$ is also sensitive to the driving force, which is in accordance with Dykman's theoretical descriptions \citep{dykman1979theory}.

In theoretical analysis of stochastic switching in Duffing oscillators, it has been predicted that both of $E_a$ and $\Gamma_0$ rely on several experimentally accessible parameters, such as driving force $f_d$, and frequency detuning between the driving frequency $\Omega_d$ and the bifurcation point $\Omega_{bd}$ \citep{dykman1979theory, defoort2015scaling}. Therefore, we make a quantitative comparison between our experimental results and the theoretical expressions, the barrier height $E_a \propto  f_d^2 (\Omega_{bd} - \Omega_d)^{\frac{3}{2}}/(\Omega_{bd} - \Omega_m)^{\frac{5}{2}}/ \gamma_m$ and the switching rate $\Gamma_0 =  \vert \Omega_{bd} - \Omega_d \vert ^{\frac{1}{2}} (\Omega_{bd} - \Omega_m)^{\frac{1}{2}}/(2\pi)$ \citep{defoort2015scaling}. From slopes of two curves shown in the Fig. \ref{sch:switchingRateTemp} (b), we have obtained  $E_a(f_2) / E_a(f_1) \approx $ 19.4, in accordance with calculation results of 18.2. However, for the switching rate $\Gamma_0$, the calculation results are almost $\sim$ 5 times higher than our measurement results and give $\Gamma_0(f_2) /\Gamma_0(f_1) \approx $ 2, higher than our measurement results, $\approx $1.3. Although it is difficult to have a quantitative agreement in comparisons, measurement results exhibit qualitative accordance with theoretical descriptions. 

\begin{figure}
  \includegraphics[width=0.48\textwidth]{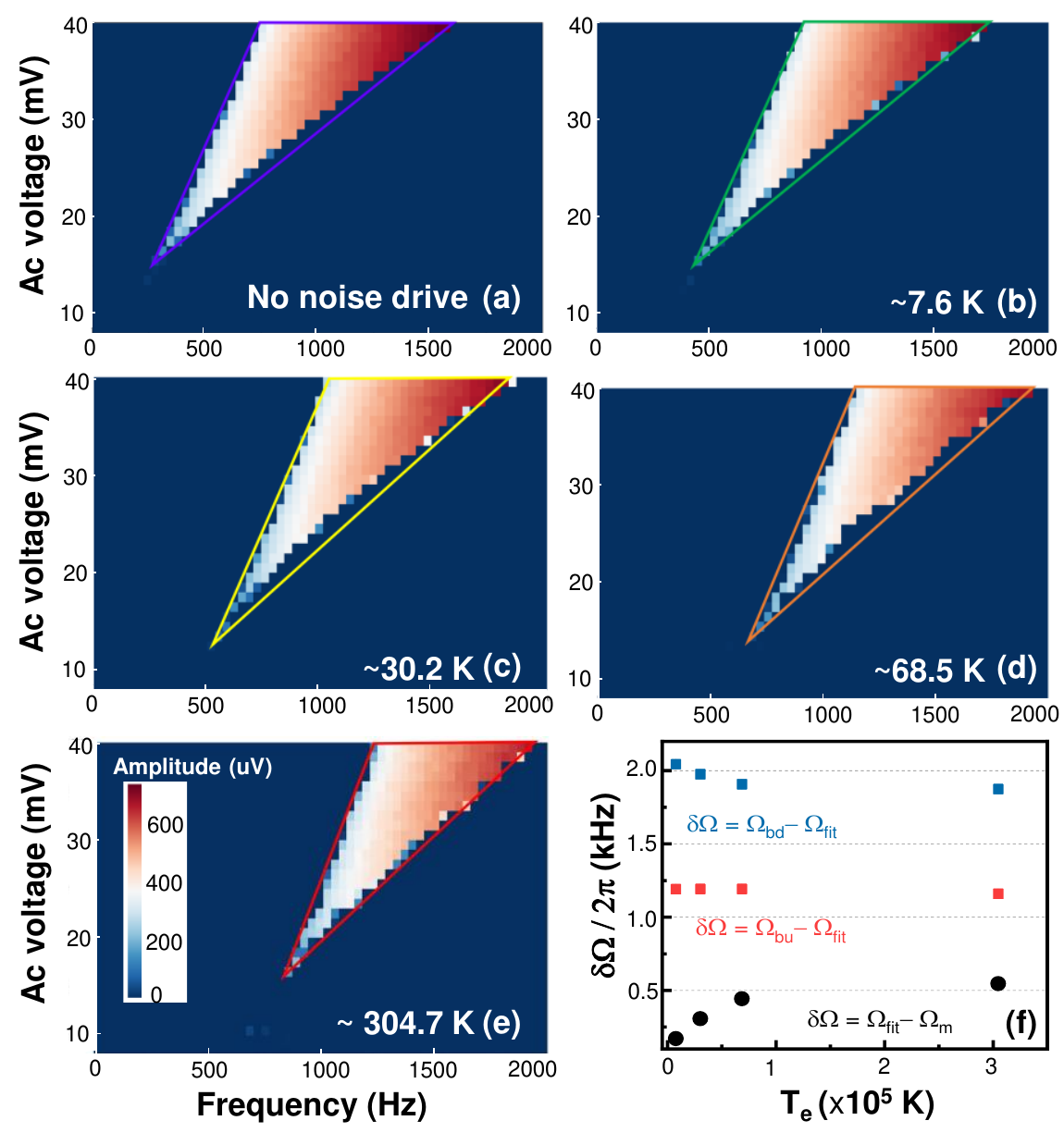}
  \caption{The measurement is performed at a fixed $dc$ voltage $V_{dc} = 2V$. (a)-(e) Hysteresis of 2D (two dimensional) plots corresponding to different noise drive power, in the form of the effective mode temperature. Each hysteresis curve is obtained from both forward and backward frequency sweeps at a fixed $V_{ac}$ voltage. All 2D plots share the same color bar of the (e). Here, the sweeping frequency is normalized by the starting point. (f) $\delta \Omega$ as a function of mechanical mode temperature, where $\delta \Omega$ = $\Omega_{bd} - \Omega_{m}$ for the forward and = $\Omega_{bu} - \Omega_{m}$ for backward sweeps respectively, measured at $V_{ac}$ = 40 mV. Here, we choose $\Omega_m$ as the reference frequency when there is no extra white noise.}
  \label{sch:hysteresis}
\end{figure}

Besides of the noise assisted random jumps in bistable states, white noise induced stochastic force affecting on mechanical properties in nonlinear region are also deserved to be investigated. Therefore, we track the variations of hysteresis window in a fixed sweeping range for both the frequency and the $V_{ac}$, when increase the amplitude of stochastic force in the form of the effective temperature $T_e$. Figure \ref{sch:hysteresis} (a)-(e) show the 2D plots of hysterics regime, corresponding to different $T_e$. Each hysteric region is obtained by subtracting the data of the backward frequency sweep from the forward one.  With increasing stochastic force, it is clear that the hysteresis window shifts towards to the higher frequency region. Here, we choose the hysteresis region, which consisting of forward and backward frequency sweeps and are measured at the $V_{ac}$ = 40 mV, for further analysis. By fitting each nonlinear curve, the resonance frequency $\Omega_{fit}$ is obtained and  we find that it shifts towards the higher frequencies as $T_e$ increases. Moreover, variations of bifurcation points as a function of $T_e$ have also been observed through definitions of  $\delta\Omega = \Omega_{bd} -\Omega_{fit}$ and $\delta\Omega =\Omega_{bu} -\Omega_{fit}$. These analyses are shown in Fig. \ref{sch:hysteresis}(f). 

Compared with $\Omega_{bu}$, the $\Omega_{bd}$ seems to be more sensitive to stochastic forces. For a Duffing oscillator, driven by a periodic force $f_d(t)$, it is well-known that the bifurcation point $\Omega_{bd}$ (for the case of spring hardening effect) can be generally described by $\Omega_{bd}$ = $\Omega_m ( 1+ \frac{3\alpha}{8k} x^2_{max})$ \citep{lifshitz2008nonlinear}. It depends on the Duffing parameter $\alpha$,  the maximum mechanical displacement $x_{max}$, and the spring constant $k$. From Fig. \ref{sch:hysteresis} (b)-(e), we did not observe clearly variations in amplitudes at $\Omega_{bd}$. Besides, for circular membrane, Duffing (hardening) nonlinear coefficient $\alpha$, is mainly dominated by geometry \citep{cattiaux2020geometrical}. Therefore, the displacements of the hysteresis window can be attributed to the modulations of spring constant by stochastic forces. Besides observations of hysteresis windows shifting, we also find that the width of hysteresis window starts to be reduced as the amplitude of stochastic force increases. Similar phenomenon has been reported in previous studies of doubly-clamped beams with  \citep{aldridge2005noise} and micro-cantilever \citep{venstra2013stochastic} increasing amplitude of stochastic force, which also has been attributed to stochastic forces affecting the spring constant of the mechanical resonator. However, compared with these previous reports, we did not observe the hysteresis quenched even though the membrane is artificially heated at same orders of the $T_e$. We suppose that it requires the much higher value of the $f_n$ to quench hysteresis regime. Because this high stressed silicon nitride membrane, with a fully clamped scheme, has the higher spring constant $k \sim$ 90 N/m \citep{zhou2021high}. 

\section{Conclusion}
In conclusion, we have investigated the effect of stochastic forces on the Duffing nonlinear behaviors of silicon nitride membrane nanoelectromechanical resonators, through room temperature measurement. In the hysteresis region, the probability of occupying a state exhibits high sensitivity to the driving frequency, demonstrating the potential to explore nonlinear-based sensing. The stochastic switching of the membrane, corresponding to different mechanical modes, has been found to follow Kramer's law. Our measurements are in qualitative agreement with the theoretical descriptions  of the maximum switching rate $\Gamma_0$ and activation energy $E_a$ influenced by the experimental setting of the driving force and the width of the hysteresis window. Besides, squeeze of Duffing hysteresis window and shifts of bifurcation points have also been observed when increasing the effective temperature $T_e$ and been attributed to stochastic forces influencing the spring constant of the membrane. Although the typical values of the $\Gamma_0$ obtained in this experiment is in the orders of tens Hz, these studies pave way for explorations of stochastic switching based novel functions beyond sensing, such as random number generators, stochastic computing, and logic gates \citep{vodenicarevic2017low, guerra2010noise, gaba2013stochastic, zahari2020analogue}. 

\begin{acknowledgments}
We would like to acknowledge financial support from STaRS-MOC Project No. 181386 the Region Hauts-de-France, from ISITE-MOST Project No. 201050, and the ANR grant MORETOME No.262047. This work was partly supported by the French Renatech network. 
\end{acknowledgments}


%
%
%
%


\nocite{*}

\bibliography{switching}

\end{document}